\documentclass[fleqn,usenatbib]{mnras}

\usepackage{newtxtext,newtxmath}

\usepackage[T1]{fontenc}
\usepackage{ae,aecompl}

\usepackage{graphicx}	
\usepackage{amsmath}	
\usepackage{amssymb}	
\usepackage{bm}		
\usepackage{pdflscape}	
\usepackage{color}
\usepackage{hyperref} 
\usepackage{lscape}

\title[Constraints on ultra-low-frequency GWs II]{Constraints on ultra-low-frequency gravitational waves with statistics of pulsar spin-down rates II: Mann-Whitney~U test}

\author[H. Kumamoto et al.]{
H. Kumamoto$^{1,2}$\thanks{E-mail: hiroki$\_$kumamoto@kumadai.jp},
S. Hisano$^{1}$,
and
K. Takahashi$^{1,3}$
\\
$^{1}$Kumamoto University, Graduate School of Science and Technology, 2-39-1 Kurokami, Chuo-ku, Kumamoto 860-8555, Japan\\
$^{2}$CSIRO Astronomy and Space Science, Australia Telescope National Facility, PO Box 76, Epping, NSW 1710, Australia\\
$^{3}$Kumamoto University, International Research Organization for Advanced Science and Technology, 2-39-1 Kurokami, Chuo-ku, Kumamoto 860-8555, Japan\\
}

\date{Accepted XXX. Received YYY: in original form ZZZ}

\pubyear{2020}

\begin{document}
\label{firstpage}
\pagerange{\pageref{firstpage}--\pageref{lastpage}}
\maketitle

\begin{abstract}
We investigate gravitational waves with sub-nanoHz frequencies ($10^{-11}$~Hz $\lesssim f_{\rm GW} \lesssim 10^{-9}$~Hz) from the spatial distribution of the spin-down rates of milli-second pulsars. As we suggested in \citet{yon18}, gravitational waves from a single source induces the bias in the observed spin-down rates of pulsars depending on the relative direction between the source and pulsar. To improve the constraints on the time derivative of gravitational-wave amplitude obtained in our previous work \citep{kum19}, we adopt a more sophisticated statistical method called the Mann-Whitney~U test. Applying our method to the ATNF pulsar catalogue, we first found that the current data set is consistent with no GW signal from any direction in the sky. Then, we estimate the effective angular resolution of our method to be $(66~{\rm deg})^2$ by studying the probability distribution of the test statistic. Finally, we investigate gravitational-wave signal from the Galactic Centre and M87 and, comparing simulated mock data sets with the real pulsar data, we obtain the upper bounds on the time derivative as $\dot{h}_{\rm GC} < 8.9 \times 10^{-19}~{\rm s}^{-1}$ for the Galactic Centre and $\dot{h}_{\rm M87} < 3.3 \times 10^{-19}~{\rm s}^{-1}$ for M87, which are stronger than the ones obtained in \cite{kum19} by factors of 7 and 25, respectively.
\end{abstract}

\section{Introduction}
\label{sec:introduction}
Pulsar timing arrays~(PTAs) can detect nanoHz-frequency gravitational waves. Although such GWs have not been detected yet, the three major PTA groups, Parkes PTA \citep{man13,ker20}, Europian PTA \citep{mcl13,bab16} and NANOGrav \citep{agg19,fai20} are operating to detect GWs and have put constraints on the GW amplitudes. These PTA groups also cooperate as the International PTA \citep{ver16} in order to improve the sensitivity. In the near future, Square Kilometre Array~(SKA) constructed in Australia and South Africa will appear and discover about 27,000~pulsars including about 3,000~millisecond pulsars~(MSPs) \citep{kea15,kra15} and further improve the sensitivity. Thus, PTAs will greatly promote the multi-wavelength gravitational-wave astronomy.

The nanoHz-frequency GWs are radiated from supermassive black hole~(SMBH) binaries in galactic cores. The frequency of GWs $f_{\rm GW}$ is determined by the separation $a$ and reduced mass $\mu$ of the binary and typically $f_{\rm GW} \sim 10^{-8}~{\rm Hz}$ for $a \sim 10^{-2}~{\rm pc}$ and $\mu = 2.5 \times 10^{8}~{\rm M}_{\odot}$. On the other hand, the frequency of observable GWs by the PTA is determined by the observational time-span and cadence. The practical range is typically $10^{-9} \lesssim {\it f}_{\rm GW} \lesssim 10^{-6}$~Hz. This frequency range corresponds to the late stage of the binary evolution.

We need to detect sub-nanoHz frequency GWs to probe the earlier phase of the binary evolution, especially to challenge the final parsec problem \citep{mil03,ryu18}. For stochastic GW backgrounds with sub-nanoHz frequencies, several detection methods have been proposed so far \citet{ber83,kop97,psh10}. \citet{ber83} estimated the contribution from stochastic GW backgrounds to the timing noise of pulsars and suggested that they can be distinguished from intrinsic irregularities of pulsars by searching for correlations between the timing noise of different pulsars. \citet{kop97}, by using binary pulsars, placed a limit on the energy density of GW backgrounds as $\Omega_{\rm GW} h^2 \lesssim 2.7 \times 10^{-4}$ in the frequency range of $1.1 \times 10^{-11}$~ Hz < ${\it f}_{\rm GW}$ < $4.5 \times 10^{-9}$~Hz, where $h$ is the Hubble constant. In \citet{psh10}, a new method to explore GW backgrounds in the frequency range from $10^{-12}$ to $10^{-8}$~Hz is suggested. This method is based on the precise measurements of pulsar periods. The second derivative of the periods from a number of pulsars gave a constrain of $\Omega_{\rm GW} h^2 \lesssim 10^{-6}$. For more previous works, see, e.g., \citet{ior14}.

In \citet{yon16}, we proposed a new method to detect sub-nanoHz GWs from a single source with the statistics of spin-down rates of MSPs. This work was motivated by the possible existence of the second SMBH(s) in M87 indicated by the significant displacement of the AGN from the luminous centre of M87 \citep{bat10}. As we describe in Section\,\ref{sec:principle}, sub-nanoHz GWs induce a bias on the observed spin-down rates of MSPs depending on the relative direction between the GW source and a pulsar. According to the sign of the bias factor, the celestial sphere is divided into two regions: one with a positive bias and another with a negative bias. Then, GWs can be probed by measuring the statistical difference in the distribution of spin-down rates between two pulsar groups from the positive and negative bias. In \citet{yon18}, we gave a rough estimate of the potential sensitivity using a simple model of the pulsar locational distribution. Here, we used the skewness as a statistical quantity to characterize the distribution of spin-down rates and the skewness difference between the two groups was considered to be a measure of GW signals. In \citet{his19}, we adopted a realistic model of the pulsar spatial distribution within Galaxy to improve the prediction of the sensitivity. Then, in \citet{kum19}, we derived upper bounds on the time derivative of the GW amplitudes from two possible GW sources, the Galactic Centre and M87.

In this study, we attempt to improve the constraints given in \citet{kum19} by adopting a more sophisticated statistical method named Mann-Whitney~U test. In Section\,\ref{sec:principle}, we review the basic idea for the detection of sub-nanoHz frequency GWs from the statistics of the spin-down rates. We describe the details of the pulsar catalogue we employ in this work in Section\,\ref{sec:catalog} and the Mann-Whitney~U test is introduced in Section\,\ref{sec:mannwhitney}. Our main results, constraints on the time derivative of GW amplitudes and the estimation of the effective angular resolution of our method, are presented in Section\,\ref{sec:results}. We give discussions and interpretations of the results in Section\,\ref{sec:discussions} and the summary is given in Section\,\ref{sec:conclusions}.

\section{Detection Principle}
\label{sec:principle}

Let us begin by briefly describing the detection method of ultra-low-frequency GWs following \citet{yon16}. Timing residuals of a pulsar induced by GWs are given by \citet{det79} as,
\begin{eqnarray}
r_{\rm GW}(t) &=& \sum_{A = + , \times} F^A(\hat{\pmb \Omega}, \hat{\pmb p}) \int^t \Delta h_A (t^{\prime}, \hat{\pmb \Omega}, \theta) dt^{\prime},
\label{e1}
\end{eqnarray}
where $\hat{\pmb p}$ and $\theta$ are the direction of the pulsar and the GW polarization angle. Here, we note that $\hat{\pmb \Omega}$ is the propagation direction of the GW, not the direction of GW source. The antenna beam pattern~$F^A(\hat{\pmb \Omega},\hat{\pmb p})$ describing the geometric factor is given by \citet{anh09}, 
\begin{eqnarray}
F^{A}(\hat{\pmb \Omega},\hat{\pmb p}) &=& \frac{1}{2}~\frac{\hat{p}^i \hat{p}^j}{(1+{\hat{\pmb \Omega}}~{\pmb \cdot}~{\hat{\pmb p}})}e^{A}_{ij}(\hat{\pmb \Omega}),
\label{e2}
\end{eqnarray}
where $e^{A}_{ij}(\hat{\pmb \Omega})$ $(A = + ,\times)$ are the GW polarization tensors written by 
\begin{eqnarray}
e^{+}_{ij}(\hat{\pmb \Omega}) &=& {\hat{{m}}_i}{\hat{{m}}_j} - {\hat{{n}}_i}{\hat{{n}}_j}, 
\label{e3}\\
e^{\times}_{ij}(\hat{\pmb \Omega}) &=& {\hat{{m}}_i}{\hat{{n}}_j} + {\hat{{n}}_i}{\hat{{m}}_j},
\label{e4}
\end{eqnarray}
where $\hat{\pmb m}$ and $\hat{\pmb n}$ are the polarization and orthonormal basis unit vectors to the propagation direction~$\hat{\pmb \Omega}$ of the GW. The elements of basis vectors can be written as, 
\begin{eqnarray}
\hat{\pmb \Omega} &=&(-{\rm cos \psi}~{\rm cos \phi}, -{\rm cos \psi}~{\rm sin \phi}, -{\rm sin \psi})
\label{e5}\\
\hat{\pmb m} &=& ({\rm sin \phi}, -{\rm cos \phi}, 0),
\label{e6}\\
\hat{\pmb n} &=& ({\rm sin \psi}~{\rm cos \phi}, {\rm sin \psi}~{\rm sin \phi}, -{\rm cos \psi}),
\label{e7}
\end{eqnarray}
with the assumption of a single GW source at the position of $(~{\rm RA},~{\rm DEC}~)=(\phi, \psi)$.

In Eq.\,(\ref{e1}), the difference in the metric perturbation between the earth and pulsar $\Delta h_A(t^{\prime},\hat{\pmb \Omega},\theta)$ is given by 
\begin{eqnarray}
\Delta h_A (t^{\prime}, \hat{\pmb \Omega}, {\theta}) &=& h_A (t, \hat{\pmb \Omega}, {\theta}) - h_A (t_p, \hat{\pmb \Omega}, \theta),
\label{e8}
\end{eqnarray}
where $t_p = t - \tau$ and $\tau =  L/c (1+{\hat{\pmb \Omega}}~{\pmb \cdot}~{\hat{\pmb p}})$ is the pulse propagation time from the pulsar at the distance $L$ to the earth. In the right-hand side, the first and second terms are called the ``earth term'' and ``pulsar term'', respectively.

In this work, we focus on the GWs with periods much longer than the observational time span, which is typically 10~years. Then the effect of GWs with much longer periods can be approximated to be linear function in time-domain. While the pulsar term has been often neglected in the literature, we investigated its importance quantitatively through Monte Carlo simulations in \citet{his19}. It was found that the pulsar term behaves as a random noise with zero average when the GW wavelength is shorter than or comparable to the typical pulsar distance ($\sim 1~{\rm kpc}$), in other words, the GW period is smaller than $O(1,000)~{\rm years}$. In this case, the effects of the pulsar term become statistically smaller for a large number of pulsar samples. On the other hand, when the GW wavelength is larger than the typical pulsar distance, it cannot be treated as a noise and a careful treatment is necessary.

Thus, in this paper, we consider only GW frequencies of $O(100)-O(1,000)~{\rm years}$ ($f_{\rm GW} \gtrsim 10^{-11}~{\rm Hz}$) and neglect the pulsar term as we did in our previous work \citet{kum19}. For such GWs, we can write Eq.\,(\ref{e8}) as
\begin{eqnarray}
\Delta h_A (t^{\prime},\hat{\pmb \Omega},\theta) &\simeq& \dot{h}_A(\hat{\pmb \Omega},\theta) t.
\label{e9}
\end{eqnarray}
Replacing Eq.\,(\ref{e1}) with Eq.\,(\ref{e9}) simplifies the integration in Eq.\,(\ref{e1}). We obtain the timing residuals induced by ultra-low-frequency GWs as,
\begin{eqnarray}
r_{\rm GW}(t) &=& \frac{1}{2}\sum_{A = + , \times} F^A(\hat{\pmb \Omega}, \hat{\pmb p}) \dot{h}_A(\hat{\pmb \Omega},\theta)t^2.
\label{e10}
\end{eqnarray}
On the other hand, the timing residual induced by the particular spin-down is written as
\begin{eqnarray}
r_{\rm spin}(t) &=& \frac{1}{2}\frac{\dot{P}}{P}~t^2,
\label{e11}
\end{eqnarray}
where, $P$ and $\dot{P}$ are the pulse period and its time derivative, respectively. Thus, both types of the timing residual have the same time dependence. Therefore, in the presence of ultra-low-frequency GWs, the observed spin-down rate is biased as
\begin{eqnarray}
\frac{\dot{P}_{\rm obs}}{P} &=& \frac{\dot{P}_{\rm 0}}{P} + \alpha(\hat{\pmb \Omega},\hat{\pmb p},\theta),
\label{e12}
\end{eqnarray}
where $\dot{P}_{\rm obs}$ and $\dot{P}_{\rm 0}$ are the observed and intrinsic spin-down rates, respectively. Here, the bias factor, $\alpha(\hat{\pmb \Omega},\hat{\pmb p},\theta)$ is written as,
\begin{eqnarray}
\alpha(\hat{\pmb \Omega},\hat{\pmb p},\theta) &=& \sum_{A={+},{\times}} F^{A}(\hat{\pmb \Omega},\hat{\pmb p}) \dot{h}_A(\hat{\pmb \Omega},\theta).
\label{e13}
\end{eqnarray}

\begin{figure}
\begin{center}
\includegraphics[width=8cm]{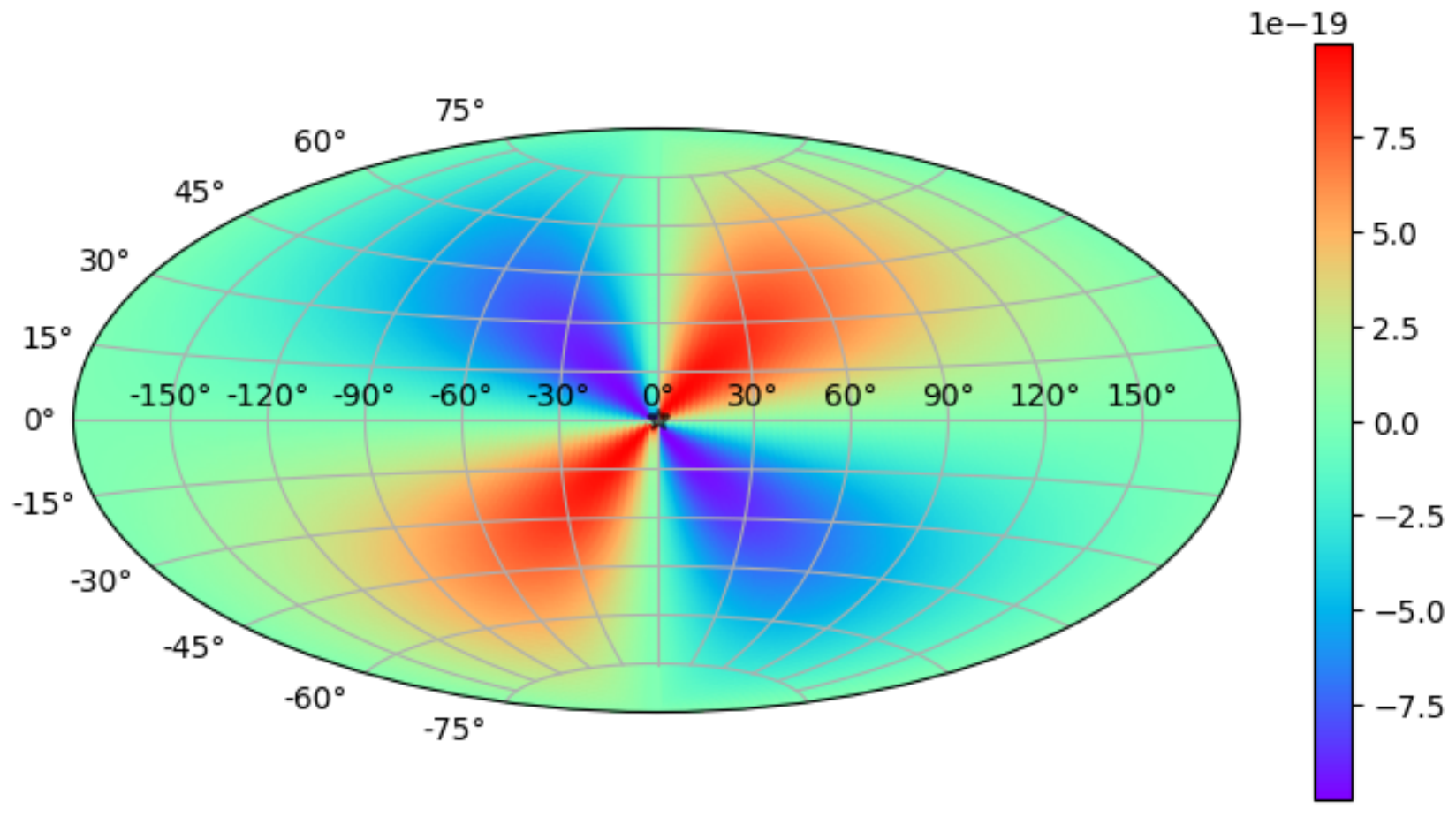}
\end{center}
\caption{The spatial pattern of the bias factor in the sky for $\dot{h}_+$ = $10^{-18}~{\rm{s}}^{-1}$. The GW source is placed at the centre of the sky in equatorial coordinates and the polarisation angle is set to $\theta = 0~{\rm deg}$.}
\label{fig:alpha}
\end{figure}

The bias factor has the quadrupole spatial pattern centred at the GW source position and the celestial sphere is divided into two areas which have positive and negative values of the bias factor $\alpha(\hat{\pmb \Omega},\hat{\pmb p},\theta)$, as shown in Fig.\,\ref{fig:alpha}. Therefore, a statistical difference in the observed spin-down rates is induced between two pulsar groups from positive and negative regions. In our previous work \citep{kum19}, the skewness of the spin-down rate distribution was considered and the skewness difference between the two regions was used to quantify the statistical difference. 

Finally, we define the two polarisation modes by,
\begin{eqnarray}
\dot{h}_+(\hat{\pmb \Omega},\theta) &=& \dot{h}(\hat{\pmb \Omega}) \cos{2\theta}, 
\label{e14}\\
\dot{h}_\times(\hat{\pmb \Omega},\theta) &=& \dot{h}(\hat{\pmb \Omega}) \sin{2\theta}.
\label{e15}
\end{eqnarray}

\section{Pulsar catalogue}
\label{sec:catalog}

In this work, we obtain observed MSP data from the ATNF Pulsar Catalogue~({\sc PSRCAT}) version 1.62 \citep{man05}. It includes 274~MSPs with the measured periods shorter than 30 $\rm{msec}$ and the time derivatives. Here, we exclude 72 MSPs in globular clusters since they would be biased significantly by the gravitational potential and complicated dynamics inside the cluster. In addition, two MSPs are removed as outliers: one with a negative spin-down rate ($\dot{P}_{\rm obs}/{P} = -10^{-20.2}~[{\rm sec}^{-1}]$,~J1801$-$3210) and one with an exceptionally large spin-down rate $(\dot{P}_{\rm obs}/{P} = 10^{-11.5}~[{\rm sec}^{-1}]$,~J0537$-$6910). Thus, just 200~MSPs are used for our analysis below. However, it should be noted that our method based on Mann-Whitney~U statistic explained below is rather robust for the presence of outliers and the results would not be affected by the removal significantly.

\begin{figure}
\begin{center}
\includegraphics[width=7cm]{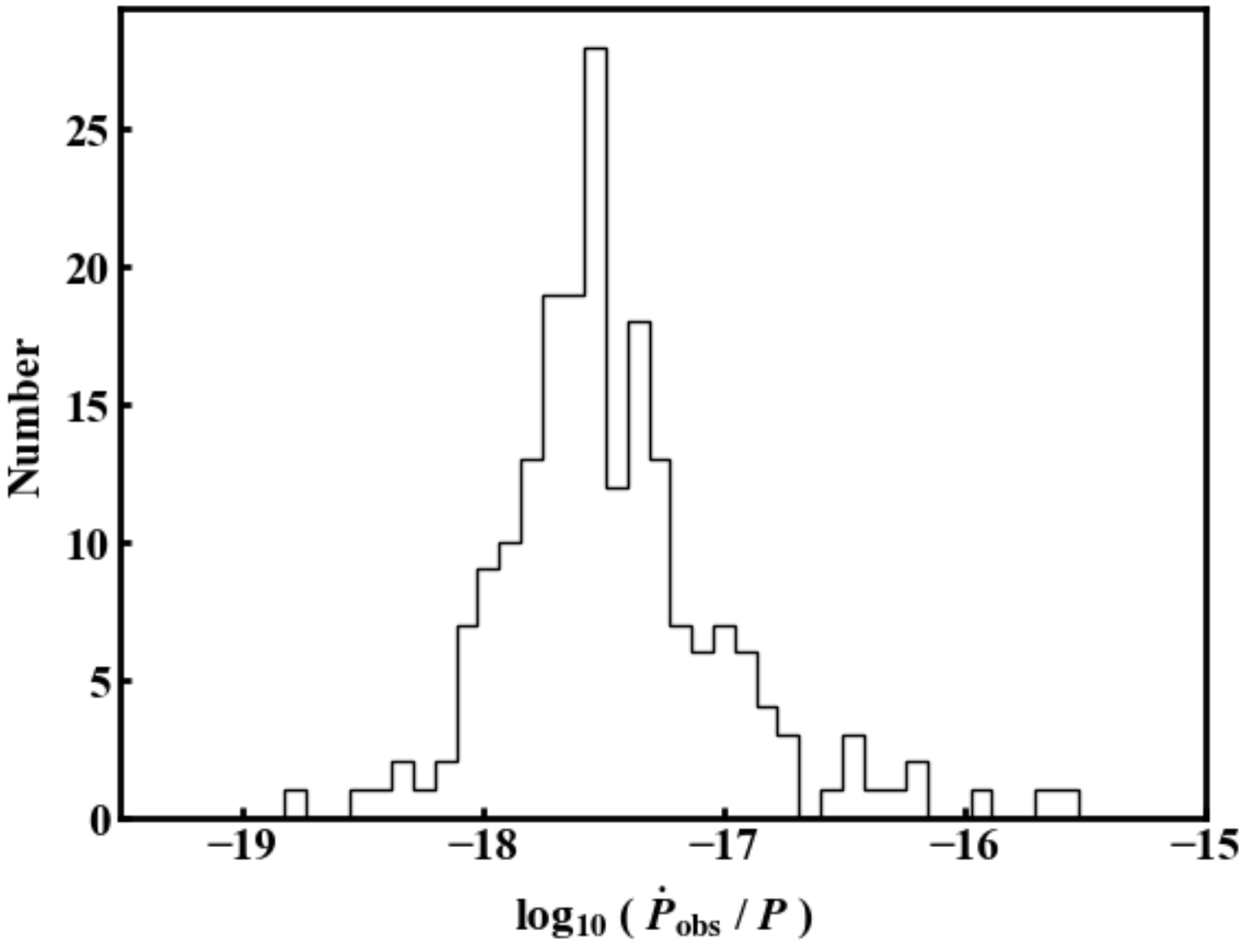}
\end{center}
\caption{The histogram of logarithmic spin-down rates ($\dot{P}_{\rm obs}/{P}$) of the 200~MSPs used in this paper.}
\label{fig:histo}
\end{figure}

\begin{figure}
\begin{center}
\includegraphics[width=7cm]{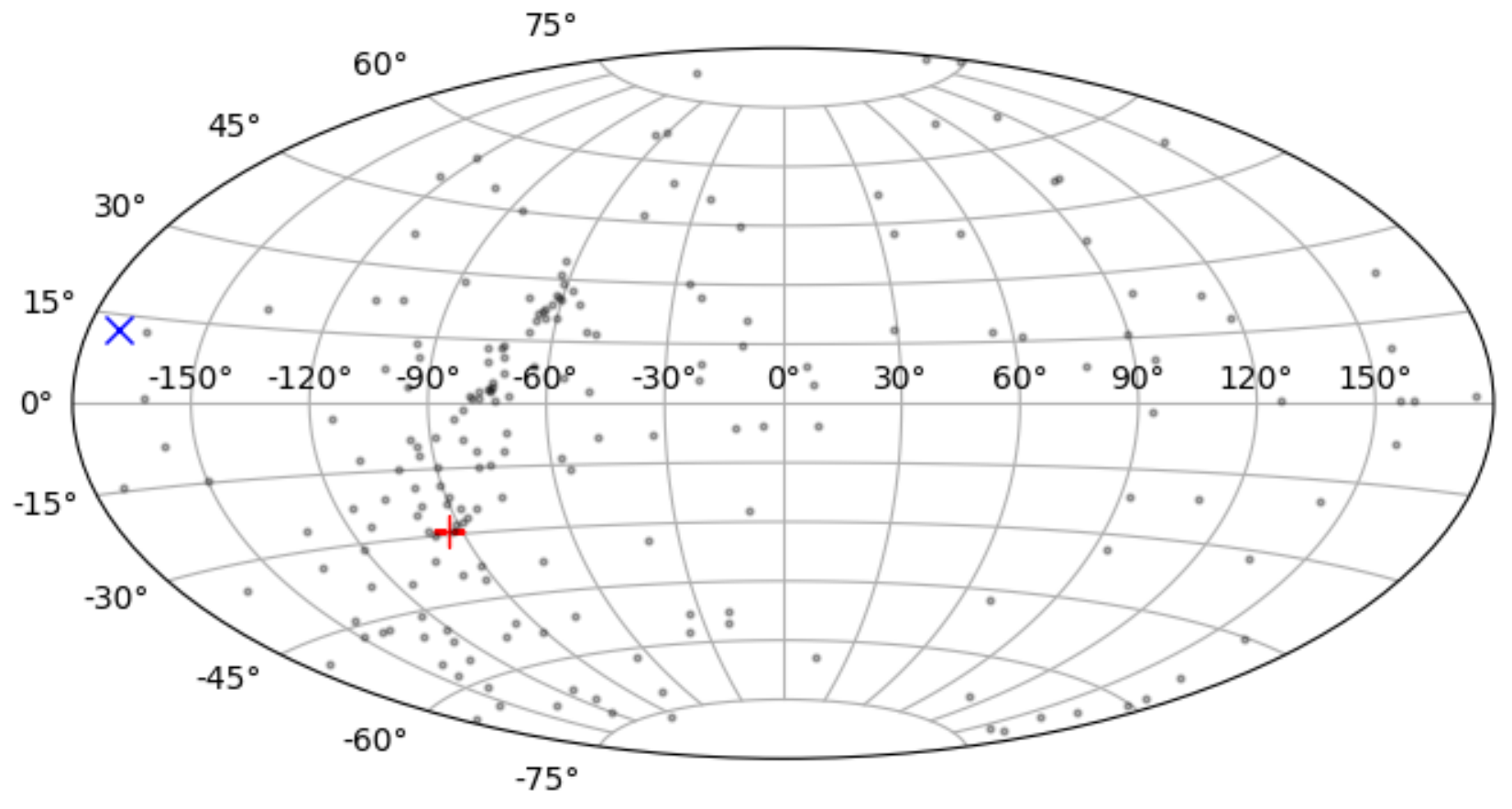}
\end{center}
\caption{Positions of the 200~MSPs in the sky in equatorial coordinates. Red "+" and blue "$\times$" show the positions of the GC and M87, respectively.}
\label{fig:position}
\end{figure}

Fig.\,\ref{fig:histo} shows the histogram of logarithmic spin-down rates ($\dot{P}_{\rm obs}/{P}$) of the observed 200~MSPs. The mean and standard deviation of this distribution are $-$17.46 and 0.47, respectively. The position of the 200~MSPs in the sky is shown in Fig.\,\ref{fig:position} in equatorial coordinates. The positions of the GC and M87 are also shown.

\section{Mann-Whitney~U Test}
\label{sec:mannwhitney}

Mann-Whitney~U test uses the ranks of samples instead of the values of spin-down rates themselves to verify if two data sets come from the same population \citep{man47}. It is non-parametric rank-based statistical test without assuming a specific form of the distribution of the data sets. Here, we apply this test to constrain the time derivative of GW amplitude. In the presence of a single GW source, the celestial sphere is divided into two groups depending on the GW polarization angle and the source position as described in Sec.~\ref{sec:principle}. Then, the spin-down rates of 200~MSPs are allocated to the two groups according to the sign of the bias factor, $\alpha(\hat{\pmb \Omega},\hat{\pmb p},\theta)$, at each pulsar. We rank the spin-down rates of all pulsars from the both groups and calculate the following
\begin{eqnarray}
{\rm U}_{B}(\hat{\pmb r},\theta) &=& {\rm n}_{+} \cdot {\rm n}_{-} + \frac{{\rm n}_{B}({\rm n}_{B}+1)}{2} - {\rm R}_{B},
\label{e16}
\end{eqnarray}
where $B~(= +, -)$ represents the group of positive and negative bias factor, respectively. Here, $\hat{\pmb r} (= -\hat{\pmb \Omega})$ is the direction of the GW source, ${\rm n}_{B}$ and ${\rm R}_{B}$ are the total sample size and rank number of each pulsar group. From ${\rm U}_{+}(\hat{\pmb r},\theta)$ and ${\rm U}_{-}(\hat{\pmb r},\theta)$, the smaller one is chosen as the Mann-Whitney~U test statistics~${\rm U}(\hat{\pmb r},\theta)$. In Mann-Whitney~U test, if the sample size of the smaller group is larger than 20, the distribution of the U test statistic is expected to follow the normal distribution. In our case, although the sample size of each group strongly depends on $\hat{\pmb r}$ and $\theta$, it is never smaller than 20. Thus, it is convenient to standardize the U test statistic using the mean value, $a = {\rm n}_{+} \cdot {\rm n}_{-} / 2$, and the standard deviation, $b = \sqrt{{\rm n}_{+} \cdot {\rm n}_{-} ({\rm n}_{+} \cdot {\rm n}_{-} +1)/12}$, as,
\begin{eqnarray}
{\rm z}(\hat{\pmb r},\theta) &=& \frac{|{\rm U}(\hat{\pmb r},\theta) - a|}{b},
\label{e17}
\end{eqnarray}
so that ${\rm z}(\hat{\pmb r},\theta)$ follows a truncated normal distribution. In order to discuss the significance of the GW signal from the direction of $\hat{\pmb r}$, we define ${\rm z_{max}}$ as,
\begin{eqnarray}
{\rm z_{max}}(\hat{\pmb r}) &\equiv& \max_{0^\circ \leq \theta \leq 180^\circ} {\rm z}(\hat{\pmb r},\theta).
\label{z_max}
\end{eqnarray}
Finally, let us define ${\rm z_{MAX}}$ by maximizing ${\rm z_{max}}(\hat{\pmb r})$ with respect to the sky position of the GW source to evaluate the overall statistical significance:
\begin{eqnarray}
{\rm z_{MAX}} &\equiv& \max_{\hat{\pmb r}} {\rm z_{max}}(\hat{\pmb r}) ~= \max_{\hat{\pmb r}, 0^\circ \leq \theta \leq 180^\circ} {\rm z}(\hat{\pmb r},\theta)
.
\label{Z_max}
\end{eqnarray}

\section{Results}
\label{sec:results}

\subsection{GW searching}
\label{subsec:gwsearching}

\begin{figure}
\begin{center}
\includegraphics[width=7cm]{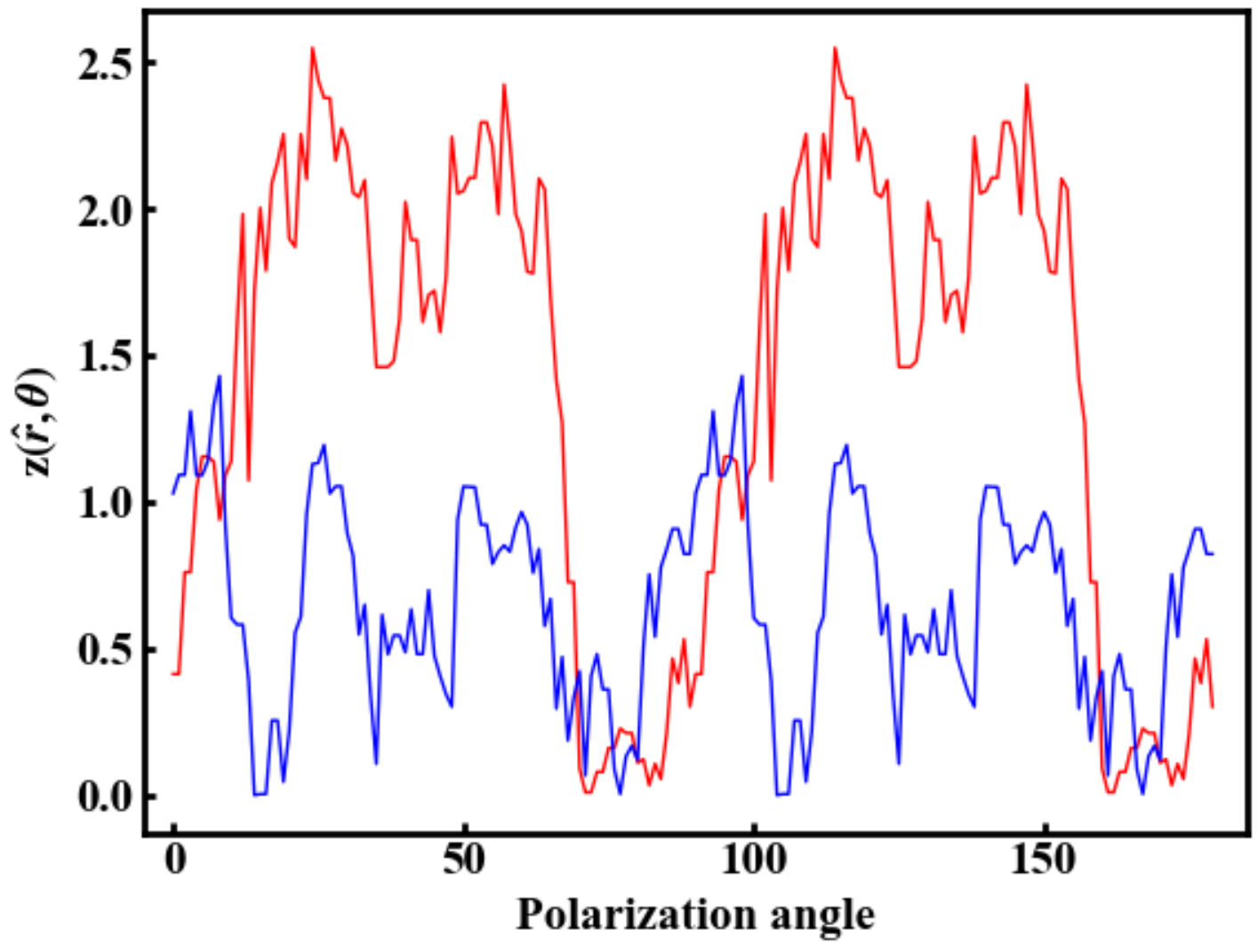}
\end{center}
\caption{The z test statistic as a function of polarization angle toward the GC and M87, ${\rm z}(\hat{\pmb r}_{\rm GC},\theta)$ and ${\rm z}(\hat{\pmb r}_{\rm M87},\theta)$. The maximum value is ${\rm z}_{\rm max}(\hat{\pmb r}_{\rm GC}) = 2.55$ and ${\rm z}_{\rm max}(\hat{\pmb r}_{\rm M87}) = 1.43$, respectively.}
\label{fig:polarization}
\end{figure}

As a demonstration, in Fig.\,\ref{fig:polarization}, we show the z test statistic as a function of the polarization angle toward the GC and M87, ${\rm z}(\hat{\pmb r}_{\rm GC},\theta)$ and ${\rm z}(\hat{\pmb r}_{\rm M87},\theta)$. As can be seen, it rapidly varies with the polarization angle and the maximum value is ${\rm z}_{\rm max}(\hat{\pmb r}_{\rm GC}) = 2.55$ and ${\rm z}_{\rm max}(\hat{\pmb r}_{\rm M87}) = 1.43$, respectively. We will use these maximum values to derive upper bounds on the time derivatives of the GW amplitudes in Section\,\ref{subsec:gw_constrain}.

\begin{figure}
\begin{center}
\includegraphics[width=8cm]{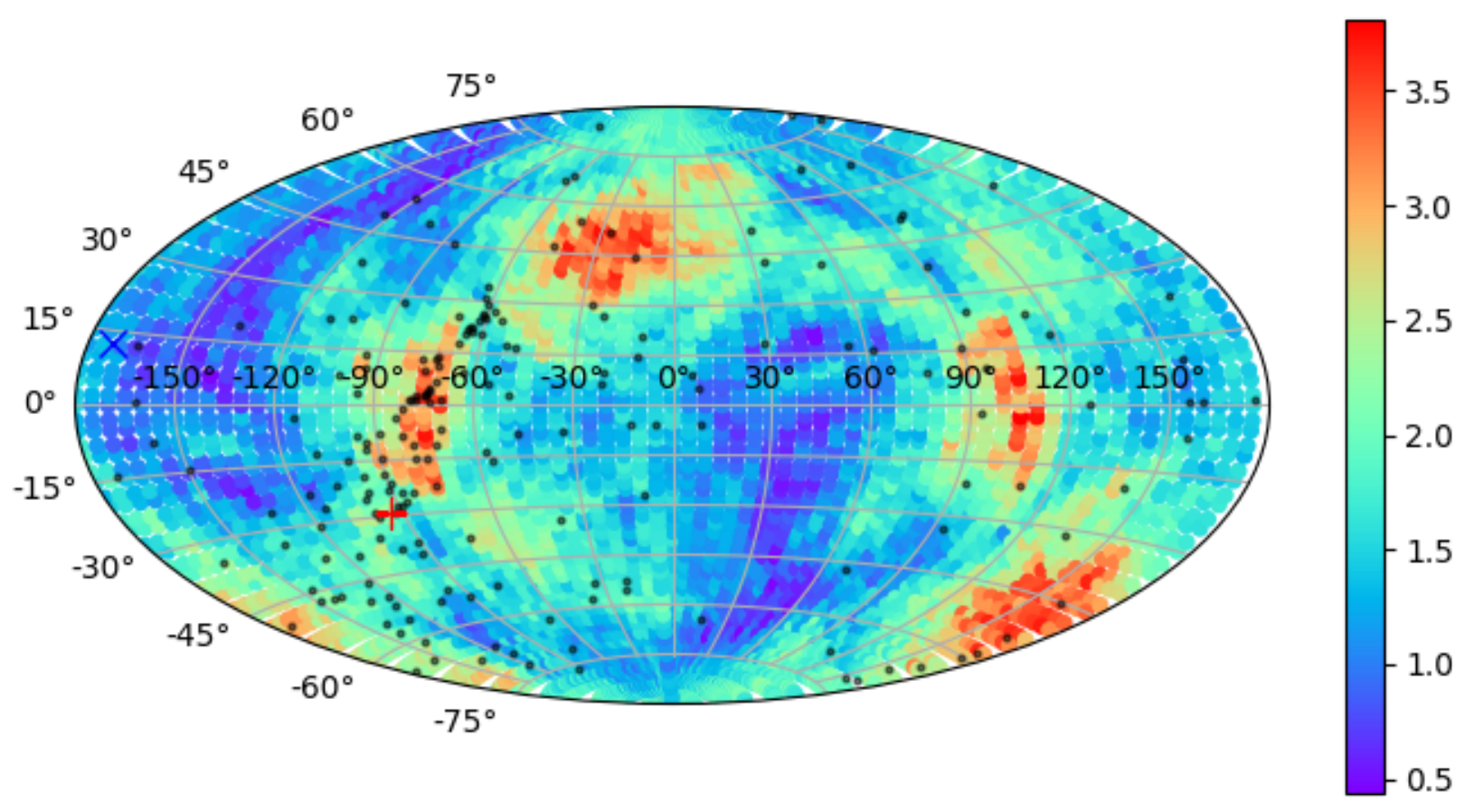}
\end{center}
\caption{The distribution of ${\rm z}_{\rm max}(\hat{\pmb r})$ in the sky. Black points represent the position of 200~MSPs. Red "+" and blue "$\times$" show the positions of the GC and M87, respectively.}
\label{fig:gwsearch}
\end{figure}

Fig.\,\ref{fig:gwsearch} shows the distribution of ${\rm z}_{\rm max}(\hat{\pmb r})$ in the sky. Here, ${\rm z}_{\rm max}(\hat{\pmb r})$ is depicted for every 5 degrees in RA and DEC, and ${\rm z}(\hat{\pmb r},\theta)$ was calculated for every 10 degrees of the polarization angle to perform the maximization. There are several hot spots where the value of ${\rm z}_{\rm max}(\hat{\pmb r})$ is relatively large ($> 3$) and the largest value is ${\rm z}_{\rm MAX} = 3.8$. The directions of this maximum value are (105 deg, $-$3 deg) and its antipode, and is away from both the GC and M87.

To evaluate the statistical significance of the value of ${\rm z}_{\rm MAX}$, we perform a series of Monte Carlo simulations. Firstly, we make a mock data set of spin-down rates ($\dot{P}_{\rm obs}/{P}$) of 200~MSPs located at the same positions as observed. Each MSP is given a value of logarithmic spin-down rate ($\log{\dot{P}/P}$) randomly following the Gaussian distribution with the same mean and standard deviation as the real data ($-$17.46 and 0.47, respectively). Then, we calculate the z test statistics in the same way as above. We perform this simulation 1,000 times and obtain the probability distribution of ${\rm z}_{\rm MAX}$.


\begin{figure}
\begin{center}
\includegraphics[width=7cm]{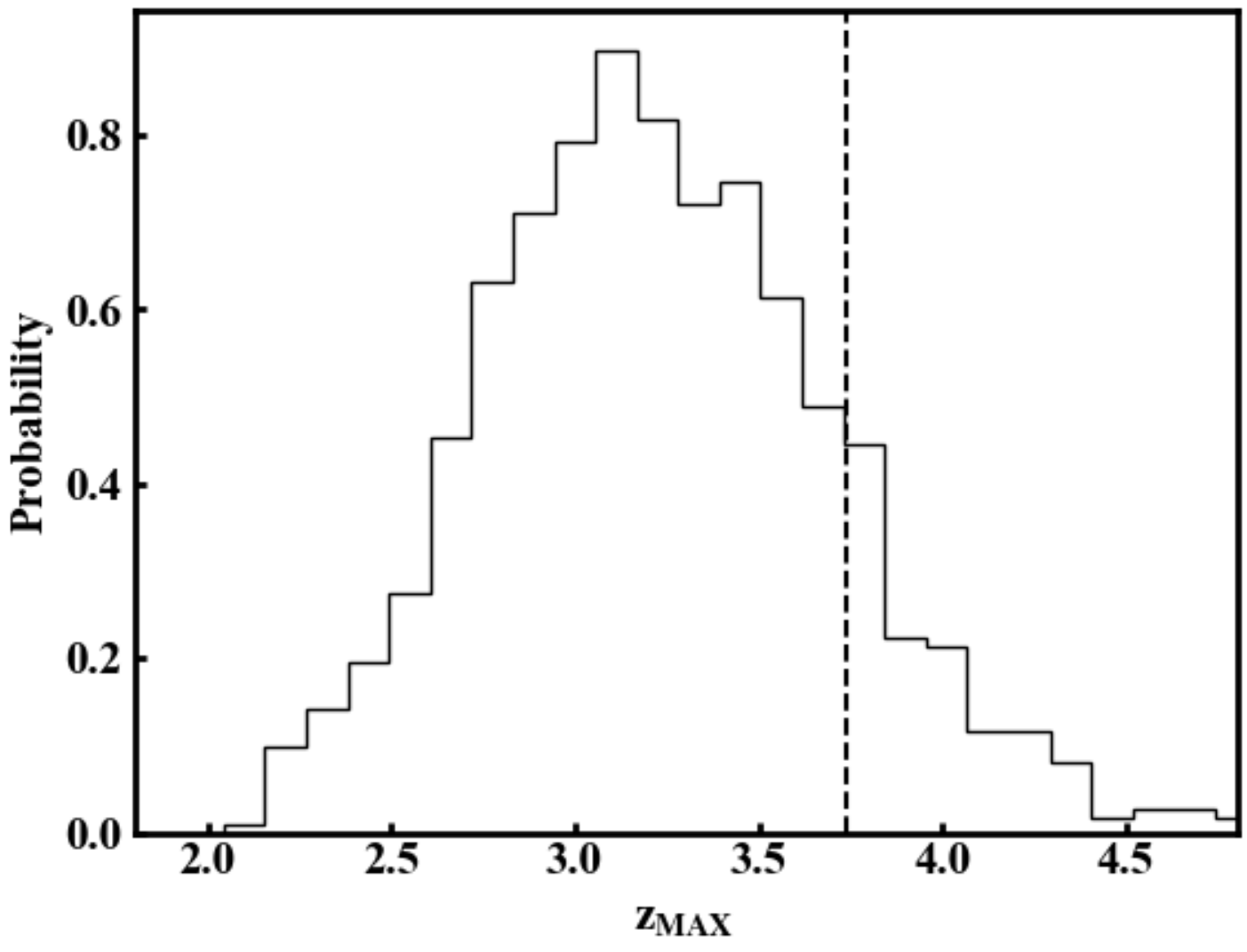}
\end{center}
\caption{The probability distribution of ${\rm z}_{\rm MAX}$ obtained from 1,000~realisations of Monte Carlo simulation without the GW injection. The vertical dashed line is ${\rm z}_{\rm MAX} = 3.8$ obtained from the real data.}
\label{fig:statistic_all}
\end{figure}

Fig.\,\ref{fig:statistic_all} shows the probability distribution of ${\rm z}_{\rm MAX}$. The distribution extends from 2.0 to 4.5 and is peaked at around $3$. The observed value of ${\rm z}_{\rm MAX} = 3.8$, indicated by the vertical line, is slightly larger than the average but is consistent with no GW signal.

\subsection{Angular resolution}
\label{subsec:angular}

In this subsection, we discuss the effective angular resolution for the GW search of our method. As we saw in Fig.\,\ref{fig:gwsearch}, although we plotted it for every 5 degrees, the spatial pattern of ${\rm z}_{\rm max}(\hat{\pmb r})$ varies with a much larger scale of about 20 degrees. This indicates that the values of ${\rm z}_{\rm max}(\hat{\pmb r})$ for adjacent pixels are not statistically independent. Thus, it is expected that the angular resolution for the GW source will be about the same order as the spacial pattern, if it is detected by our method.

\begin{figure}
\begin{center}
\includegraphics[width=7cm]{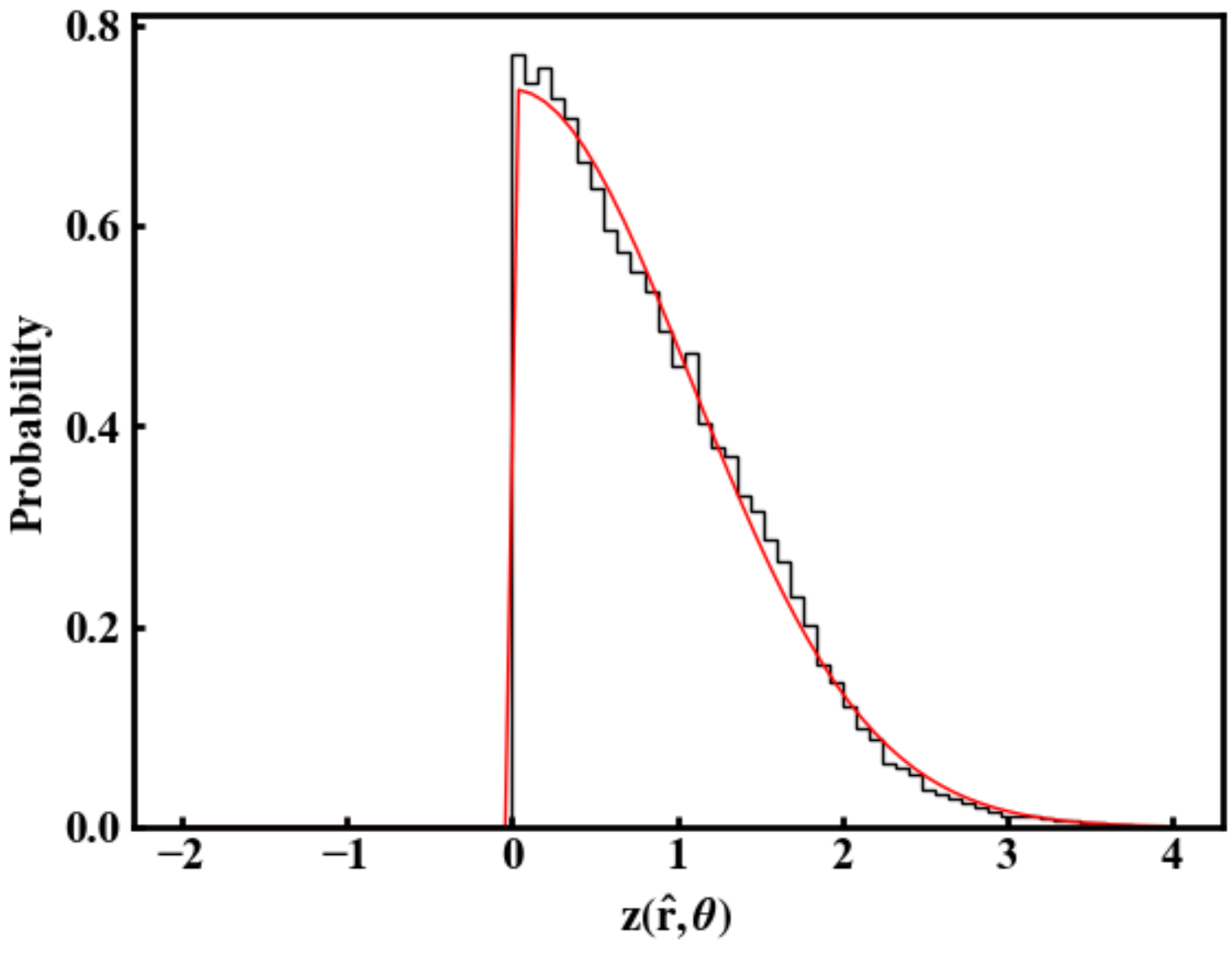}
\end{center}
\caption{The probability distribution of ${\rm z}(\hat{\pmb r},\theta)$ calculated using the z values of all positions in the sky and polarization angles of the real data. The red curve is the truncated normal distribution. See the main text for the details.}
\label{fig:zdist}
\end{figure}

The effective angular resolution can be evaluated by the statistical behaviour of the z test statistic. First, we show the probability distribution of ${\rm z}(\hat{\pmb r},\theta)$ in Fig.\,\ref{fig:zdist}. This is calculated using the z values of all positions in the sky and polarization angles of the real data. The distribution is well reproduced by the truncated normal distribution, which is the distribution of the absolute value which follows the normal distribution. The mean and standard deviation of the normal distribution (not the truncated normal distribution) are 0.0 and 1.08, respectively.


\begin{figure}
\begin{center}
\includegraphics[width=7cm]{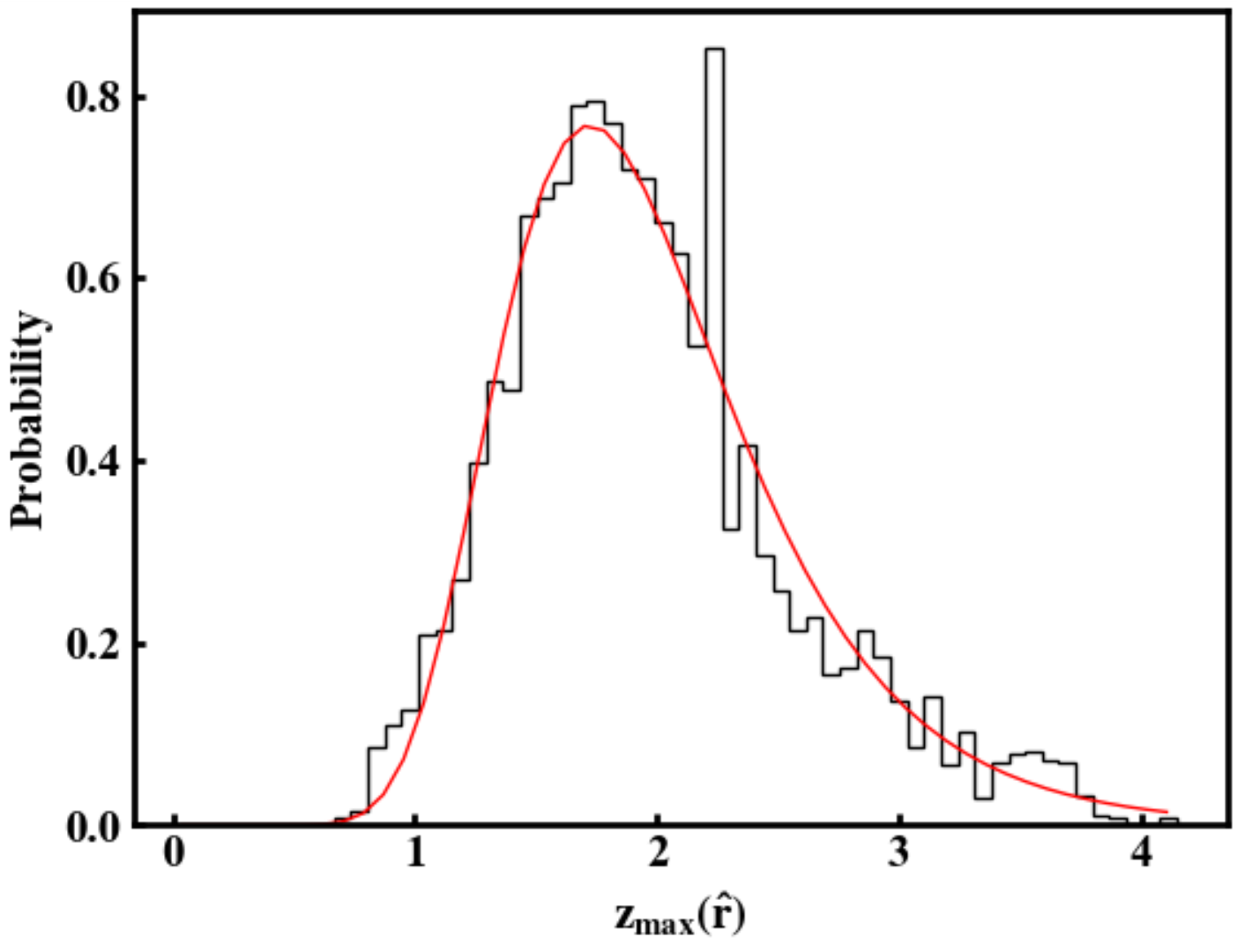}
\end{center}
\caption{The histogram of ${\rm z}_{\rm max}(\hat{\pmb r})$ shown in Fig.\,\ref{fig:gwsearch}. Red solid line represents the best-fitting Gumbel distribution with the location parameter of 1.72 and the scale parameter of 0.48.}
\label{fig:maximumdistribution}
\end{figure}

The maximum value of the z test statistic over the polarization angle, ${\rm z}_{\rm max}(\hat{\pmb r})$, follows the Gumbel distribution \citep{gum63}. It represents the distribution of the maximum value of a number of samples of various distributions and is characterized by the location parameter, $\mu$, and the scale parameter, $\beta$. The probability distribution function (PDF) for the Gumbel distributions is given as
\begin{eqnarray}
{\rm PDF}({\rm z}, \mu, \beta) &=& \exp{(-(x+{\rm e}^{-x}))}.
\label{gumbel_pdf}
\end{eqnarray}
where $x=({\rm z}-\mu)/\beta$. If we assume the z test statistic is statistically uniform in the sky, Fig.\,\ref{fig:gwsearch} can be seen as multiple realizations of the Gumbel distribution. Fig.\,\ref{fig:maximumdistribution} represents the histogram of ${\rm z}_{\rm max}(\hat{\pmb r})$ shown in Fig.\,\ref{fig:gwsearch}. In fact, it is well fitted by the Gumbel distribution with the location parameter of 1.72 and the scale parameter of 0.48, which justifies our assumption on the statistical uniformity of the z test statistic.


The two parameters of the Gumbel distribution are related to the sample size $n_s$ as
\begin{eqnarray}
\mu &=& {\rm CDF}^{-1} \left( 1-\frac{1}{n_s} \right) ,
\label{e18}\\
\beta &=& {\rm CDF}^{-1} \left( 1-\frac{1}{n_s e} \right) - \mu ,
\label{e19}
\end{eqnarray}
where CDF is the cumulative distribution function of the truncated normal distribution and $e$ is the base of the natural logarithm. From the estimated values of $\mu$ and $\beta$ above, we obtain $n_s \sim 9$. This indicates that the effective number of independent samples in Fig.\,\ref{fig:gwsearch} is $9$, which is roughly the same number of hot and cold spots there. Thus, the effective angular resolution of our method is evaluated as $40,000~{\rm deg}^2 / 9 \sim 4,400~{\rm deg}^2 \sim (66~{\rm deg})^2$.



\subsection{Constraints on the time derivative of GW amplitude}
\label{subsec:gw_constrain}

Finally, we derive constraints on the time derivative of GW amplitude focusing on two specific potential GW source candidates: GC and M87. As we saw in Section\,\ref{subsec:gwsearching}, we had the maximized z test statistics toward these directions as ${\rm z}_{\rm max}(\hat{\pmb r}_{\rm GC}) = 2.55$ and ${\rm z}_{\rm max}(\hat{\pmb r}_{\rm M87}) = 1.43$, respectively.

We evaluate the statistical significance of these values with Monte Carlo simulations similar to those given in Section\,\ref{subsec:gwsearching}. But we here inject the GW signal in order to derive constraints on the time derivative of GW amplitude. Therefore, to obtain mock observed spin-down rates, $\dot{P}_{\rm obs}/{P}$, we need to add the bias factor, $\alpha(\hat{\pmb \Omega},\hat{\pmb p},\theta)$, to the mock intrinsic spin-down rates, $\dot{P}_0/{P}$, generated randomly from the normal distribution (see Eq.\,(\ref{e12})).

\begin{figure}
\begin{center}
\includegraphics[width=7cm]{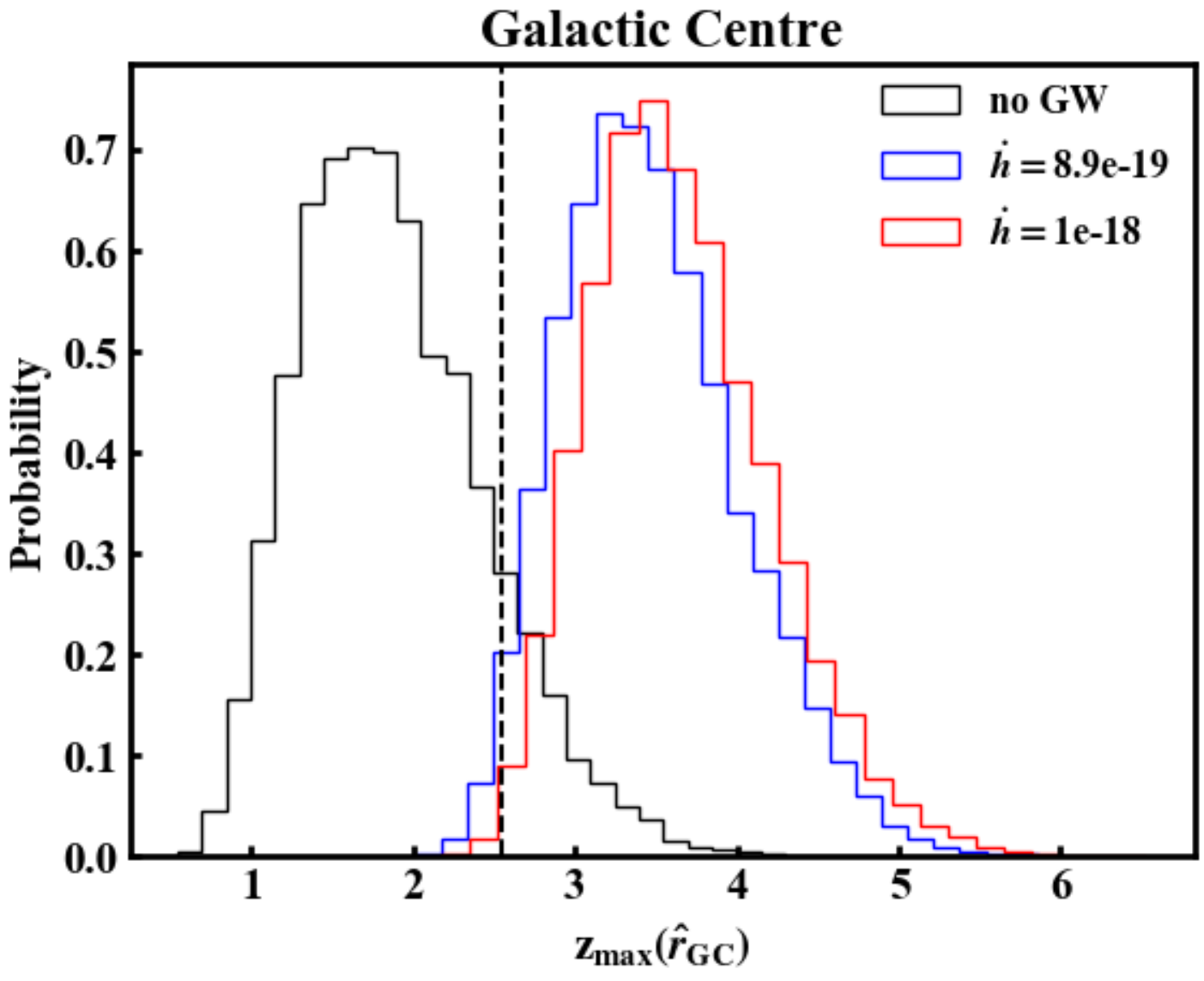}\\
\includegraphics[width=7cm]{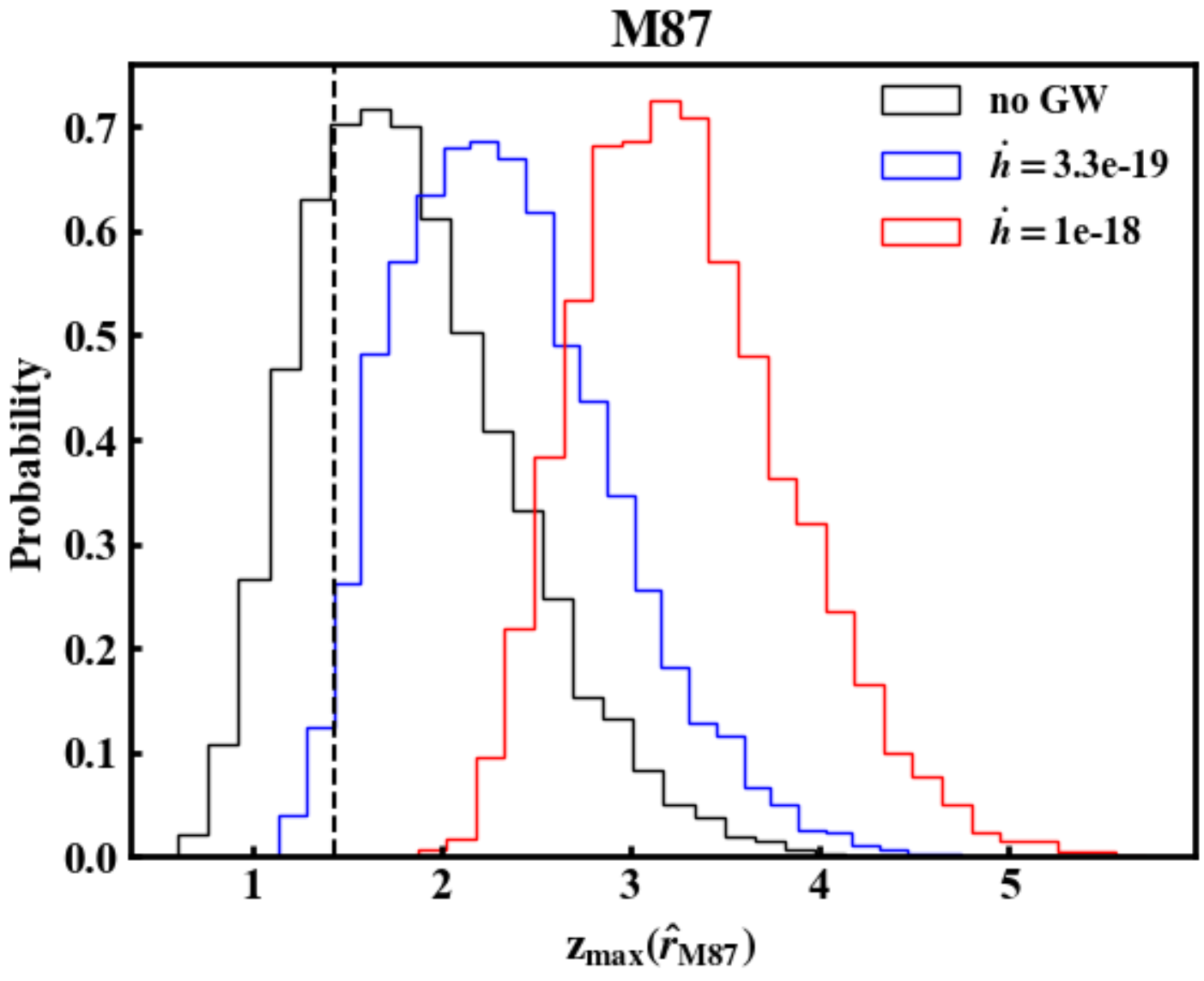}
\end{center}
\caption{The probability distribution of the maximized z test statistics ${\rm z}_{\rm max}(\hat{\pmb r})$ without the GW signal (black) and with the GW signals (red and blue). Top panel is for the GC and the red and blue lines correspond to the value of $\dot{h}$ of $10^{-18}{\rm s}^{-1}$ and $8.9 \times 10^{-19}{\rm s}^{-1}$, respectively. Bottom panel is for the M87 and the red and blue lines correspond to the value of $\dot{h}$ of $10^{-18}{\rm s}^{-1}$ and $3.3 \times 10^{-19}{\rm s}^{-1}$, respectively. The vertical dashed lines are the observed values of ${\rm z}_{\rm max}(\hat{\pmb r})$.}
\label{fig:bound_ps}
\end{figure}

Fig.\,\ref{fig:bound_ps} shows the probability distribution of ${\rm z}_{\rm max}(\hat{\pmb r}_{\rm GC})$ and ${\rm z}_{\rm max}(\hat{\pmb r}_{\rm M87})$ obtained from 10,000 realisations of simulations with and without the GW signal. As the GW signal get stronger, the probability distribution moves rightward. We regard a value of $\dot{h}$ as an upper bound, when the probability that ${\rm z}_{\rm max}(\hat{\pmb r})$ is smaller than the observed value, indicated by the vertical line, is 2\% with the assumed GW signal. We obtained upper bounds of $\dot{h}_{\rm GC} < 8.9 \times 10^{-19}~{\rm s}^{-1}$ for GC and $\dot{h}_{\rm M87} < 3.3 \times 10^{-19}~{\rm s}^{-1}$ for M87. We will discuss the implications of these upper bounds in Section\,\ref{sec:discussions}. 

\section{Discussions}
\label{sec:discussions}

First, let us compare the results with our previous work. As mentioned in Section\,\ref{sec:introduction}, in \citet{kum19}, we placed constraints on the time derivative of the GW amplitude with the skewness difference as the indicator. In consequence, we obtained $\dot{h}_{\rm GC} < 6.2 \times 10^{-18}~{\rm s}^{-1}$ for GC and $\dot{h}_{\rm M87} < 8.1 \times 10^{-18}~{\rm s}^{-1}$ for M87. Therefore, the current constraints are stronger by factors of 7 and 25 for GC and M87, respectively.

In fact, in the current analysis, we employed the ATNF pulsar catalogue version 1.62, which includes 50 more MSPs than the catalogue version 1.59 which was employed in \citet{kum19}. Thus, it is possible that the improvement of the constraints is due to the increased number of MSPs, rather than the change of the methodology. To see this, we repeated the current analysis with MSPs of the catalogue version 1.59. As a result, we obtained the maximum z statistics of ${\rm z}_{\rm max}(\hat{\pmb r}_{\rm GC}) = 2.27$ and ${\rm z}_{\rm max}(\hat{\pmb r}_{\rm M87}) = 2.02$, and the upper bounds of $\dot{h}_{\rm GC} < 8.3 \times 10^{-19}~{\rm s}^{-1}$ for the GC and $\dot{h}_{\rm M87} < 9.0 \times 10^{-19}~{\rm s}^{-1}$ for M87. The constraint for the GC is comparable to the one from the fiducial analysis. On the other hand, it is degraded by a factor of 3 for M87 but still better than the constraint in \citet{kum19} by a factor of 9. Thus, we conclude that the current method with the Mann-Whitney~U test is more effective than the previous method with the skewness difference. The comparison of the constraints is summarized in Table\,\ref{tab:hdot}.

\begin{table}
    \centering
    \caption{The comparison of the constrains on $\dot{h}$.}
	\label{tab:hdot}
\begin{tabular}{lccc}
\hline
 GW Source & This work & Kumamoto et al. & U test with \\
 & & 2019 & catalogue ver. 1.59 \\
\hline
\hline
 GC & 8.9 $\times 10^{-19}$ & 6.2 $\times 10^{-18}$ & 8.3 $\times 10^{-19}$ \\
 M87 & 3.3 $\times 10^{-19}$ & 8.1 $\times 10^{-18}$ & 9.0 $\times 10^{-19}$ \\
\hline
\end{tabular}
\end{table}

In this paper, we obtained constraints on $\dot{h}$ rather than $h$ itself. Although it is generally impossible to translate the constraints into ones on $h$, typical values can be estimated by using $\dot{h} \approx 2\pi {\it f}_{\rm GW} h$. Then, our constrains can be converted as $h_{\rm GC} \lesssim 4.5 \times 10^{-8}$~(100~years~/~${\it f}_{\rm GW}$) and $h_{\rm M87} \lesssim 1.7 \times 10^{-8}$~(100~years~/~${\it f}_{\rm GW}$), respectively. These constraints on $h$ can be further translated into upper bounds on the mass of a possible second supermassive black hole at these places. The GW amplitude~$h$ is related with the chirp mass ${\cal M}$ of the binary, the frequency $f_{\rm GW}$ of GW and the distance $L$ to the source, and described in \citet{yon18} as
\begin{eqnarray}
h = \frac{2(G{\cal M})^{5/3}(\pi f_{\rm GW})^{2/3}}{c^4 L}
\label{chirpmass}
\end{eqnarray}
where ${\cal M} = (m_1 m_2)^{3/5} / (m_1 + m_2)^{1/5}$ is the chirp mass of the binary, $m_1$ and $m_2$ are the masses of SMBHs, $L$ is the distance to the binary, $G$ is the gravitational constant and $c$ is the speed of light. For the Galactic Centre, \citet{gil17} monitored stellar orbits and obtained the current best estimates for the mass and distance of SgrA* as $m_1 = 4.3 \times 10^{6}~{\rm M}_{\odot}$ and $L = 8.3~{\rm kpc}$. On the other hand, \citet{aki19} obtained an estimate for the mass of the SMBH in M87 as $m_1 = 6.6 \times 10^{9}~{\rm M}_{\odot}$ and the distance has been estimated to be $L = 16.8~{\rm Mpc}$ \citep{bla09,can18}. Assuming a circular orbit with the orbiting period of 100~years and zero inclination (face-on), upper bounds on the second BH mass~$m_2$ are $9.6 \times 10^{14}~{\rm M}_{\odot}$ for the GC and $2.7 \times 10^{14}{\rm M}_{\odot}$ for M87. These numbers are also improved significantly compared to those obtained in \citet{kum19}, $2 \times 10^{16}~{\rm M}_{\odot}$ for the GC and $4 \times 10^{16}~{\rm M}_{\odot}$ for M87, although the astrophysical impact is still rather small.

In the SKA era, we will have 3,000~MSPs \citep{kea15,kra15} and the constraint on ultra-low-frequency GWs considered here will improve accordingly. In fact, we estimated the future constraints with the skewness difference in \citet{his19} and $\dot{h} < 2 \times 10^{-19}~{\rm s}^{-1}$ at $f_{\rm GW} = 10^{-11}$~Hz was obtained for the GC, which is stronger than the constraint obtained in \citet{kum19} by a factor of 30. It is possible that a future constraint based on the Mann-Whitney~U test statistics will improve by a similar factor. This possibility will be pursued elsewhere.

\section{Conclusions}
\label{sec:conclusions}

We constrained the time derivative of GW amplitude with sub-nanoHz frequencies ($10^{-11}$~Hz $\lesssim f_{\rm GW} \lesssim 10^{-9}$~Hz) from the spatial distribution of the spin-down rates of MSPs. As we suggested in \citet{yon18}, the GW from a single source induce the bias in the observed spin-down rates of pulsars depending on the relative direction between the GW source and pulsar. Compared with our previous studies \citep{yon18,his19,kum19}, where the skewness difference in the spin-down rate distribution was considered to detect the bias, we adopted a more sophisticated statistical method called the Mann-Whitney~U test. 

Applying our method to the ATNF Pulsar Catalogue version 1.62, we first found that the maximized value of the z test statistic obtained from the current data set is consistent with no GW signal from any direction in the sky. Then, we estimated the effective angular resolution of our method to be $(66\,{\rm deg})^2$ by studying the probability distribution of the z test statistic. Finally, comparing simulated mock data sets with the real pulsar data, the upper bounds on $\dot{h}$ were derived as $\dot{h}_{\rm GC} < 8.9 \times 10^{-19}~{\rm s}^{-1}$ for the GC and $\dot{h}_{\rm M87} < 3.3 \times 10^{-19}~{\rm s}^{-1}$ for M87, which are stronger than the ones obtained in \cite{kum19} by factors of 7 and 25, respectively. These constraints would be improved significantly with 3,000 MSPs expected to be discovered by the SKA.


\section*{Acknowledgements}

The Parkes telescope is a part of the Australia Telescope National Facility which is funded by the Commonwealth of Australia for operation as a National Facility managed by CSIRO.
The ATNF Pulsar Catalogue at \href{http://www.atnf.csiro.au/people/pulsar/psrcat/}{http://www.atnf.csiro.au/people/pulsar/psrcat/} was used for this work.
KT is partially supported by JSPS KAKENHI Grant Numbers JP15H05896, JP16H05999, and JP17H01110, and Bilateral Joint Research Projects of JSPS.
SH is supported by JSPS KAKENHI Grant Numbers JP20J20509.

\bibliographystyle{mnras}

\bsp	
\label{lastpage}
\end{document}